\documentclass[review, authoryear,11pt]{elsarticle}

\makeatletter
\def\ps@pprintTitle{%
 \let\@oddhead\@empty
 \let\@evenhead\@empty
 \def\@oddfoot{}%
 \let\@evenfoot\@oddfoot}
\makeatother

\usepackage{amssymb}
\usepackage[fleqn]{amsmath}
\usepackage{multicol}
\usepackage{enumitem}
\usepackage{lineno}
\usepackage[a4paper]{geometry}
\usepackage{natbib}
\usepackage{xcolor}

\definecolor{amber}{rgb}{0.7, 0.5, 0.0}

\newcommand\hl[1]{{\color{black}#1}}

\newcommand{\df}[0]{{\rm d}}

\newcommand*\patchAmsMathEnvironmentForLineno[1]{%
  \expandafter\let\csname old#1\expandafter\endcsname\csname #1\endcsname
  \expandafter\let\csname oldend#1\expandafter\endcsname\csname end#1\endcsname
  \renewenvironment{#1}%
     {\linenomath\csname old#1\endcsname}%
     {\csname oldend#1\endcsname\endlinenomath}}%
\newcommand*\patchBothAmsMathEnvironmentsForLineno[1]{%
  \patchAmsMathEnvironmentForLineno{#1}%
  \patchAmsMathEnvironmentForLineno{#1*}}%
\AtBeginDocument{%
\patchBothAmsMathEnvironmentsForLineno{equation}%
\patchBothAmsMathEnvironmentsForLineno{align}%
\patchBothAmsMathEnvironmentsForLineno{flalign}%
\patchBothAmsMathEnvironmentsForLineno{alignat}%
\patchBothAmsMathEnvironmentsForLineno{gather}%
\patchBothAmsMathEnvironmentsForLineno{multline}%
}

\journal{Journal of Theoretical Biology}

\begin{document}

\begin{frontmatter}

\title{Stable cycling in quasi-linkage equilibrium: fluctuating dynamics under gene
conversion and selection}

\author[address1]{Timothy W. Russell\corref{cor1}}
\ead{timothy.russell.2015@rhul.ac.uk}
\cortext[cor1]{Corresponding author}
\author[address2]{Matthew J. Russell}
\author[address1]{Francisco \'Ubeda}
\author[address1]{Vincent A.A. Jansen}

\address[address1]{School of Biological Sciences, Royal Holloway
University of London, Egham, Surrey, TW20 0EX, UK}
\address[address2]{School of Mathematical Sciences, University of Nottingham,
University Park, Nottingham, NG7 2RD, UK}

\begin{abstract}

    Genetic systems with multiple loci can have complex dynamics. For example, mean
    fitness need not always increase and stable cycling is possible. \hl{Here, we study the
    dynamics of a genetic system inspired by the molecular biology of
    recognition-dependent double strand breaks and repair as it happens in recombination
    hotspots. The model shows slow-fast dynamics in which the system converges to the
    quasi-linkage equilibrium (QLE) manifold. On this manifold, sustained cycling is
    possible as the dynamics approach a heteroclinic cycle, in which allele frequencies
    alternate between near extinction and near fixation. We find a closed-form
    approximation for the QLE manifold and use it to simplify the model. For the
    simplified model, we can analytically calculate the stability of the heteroclinic
    cycle. In the discrete-time model the cycle is always stable; in a
    continuous-time approximation, the cycle is always unstable. This demonstrates that
    complex dynamics are possible under quasi-linkage equilibrium.
}

\end{abstract}


\end{frontmatter}

\section{Introduction}
\label{sec:intro}


\hl{Genetic equilibrium, the idea that gene frequencies are the same from one generation to
the next, was the focus of early work on population genetics. The attention shifted when
it was discovered that one-locus viability models can exhibit cycling behaviour and
genetic equilibrium does not have to be achieved \citep{kimura1958change,
hadeler1975selection, asmussen1977density, cressman1988frequency}. Further
investigation showed that two-locus viability models with recombination can also
exhibit cycling behaviour \citep{akin1979geometry, hastings1981stable,
akin1982cycling, akin1983hopf, akin1987cycling}.

The discrete-time selection-recombination equations \citep{lewontin1960evolutionary,
burger2000mathematical} have provided a determinsitic model for changes in the genetic
make up of a population. Despite the fact that these equations are often used to study the
properties of stable equilibria, they are inherently nonlinear, meaning even the most
simple formulations of the equations can have complex dynamics.  Examples include limit
cycles \citep{akin1983hopf} and heteroclinic cycles \citep{haig1991genetic,
ubeda2019prdm9}.  Whether the cycles are maintained indefinitely or eventually die out
(i.e. their stability properties) is mathematically challenging and of significant
biological importance. This is the focus of the research we present here.

Many genetic processes within an interacting population of individuals can be captured by
the selection-recombination equations, as they allow for arbitrary selection regimes
defined by model-specific fitness matrices. Here, we investigate the stability of cycles
in two-locus genetic systems characterised by a specific interaction between selection,
gene conversion and crossover. This interaction corresponds to a model of the evolution of
recombination hotspots \citep{ubeda2019prdm9}. However, we re-write this model in standard
selection-recombination equations form by noticing that the effect of conversion in
\cite{ubeda2019prdm9} can be split into its effect on selection (and incorporated to the
selection component of the standard selection-recombination equation) and its effect on
formation of double heterozygotes (and incorporated into the recombination component of
the standard selection-recombination equation).  Furthermore, while the model in
\cite{ubeda2019prdm9} assumes that the values taken by the selection-recombination
parameters are constrained by their biological interdependence, here we assume that the
parameter values are independent and not limited by biological constraints. In doing so,
we allow for multiple forms of interaction between selection, conversion and crossover,
provided they produce the same equations. This formulation allow us to focus on the
mathematical properties of the generalised model.

Biologically, the processes in our model are initiated by recognition between a protein
formed by a modifier gene and a target locus, whereby the protein interacts with the
target, initiating conversion and potentially crossover \citep{ubeda2011red,
ubeda2019prdm9}.  Other than the evolution of recombination hotspots
\citep{ubeda2011red, ubeda2019prdm9}, examples of similar recognition-initiated
interactions producing sustained cycling include: the evolution of homing
endonucleases \citep{yahara2009evolutionary}, the evolution of meiotic drive
\citep{haig1991genetic}, the evolution of host-parasite interactions
\citep{sasaki2002clone} and the evolution of altruism via tag based recognition
\citep{jansen2006altruism}.

If selection is weak, stable cycling cannot occur within the two-locus
selection-recombination equations if the equilibria are hyperbolic
\citep{nagylaki1999convergence, pontz2018evolutionary}. These conditions produce dynamics
which converge to a stable equilibrium. Under weak selection, the argument by
\cite{nagylaki1999convergence} uses the existence of an invariant stable manifold which
attracts the dynamics. On this attracting manifold, the dynamics are gradient-like and
converge to equilibrium \citep{pugh1977invariant}. This manifold is known in genetics as
the \textit{quasi-linkage equilibrium} (QLE) manifold \citep{kimura1965attainment}. It is
the set of states defined by the property that linkage disequilibrium changes an order of
magnitude slower than the allele frequencies \citep{kimura1965attainment}.

In geometrical terms, this means that the dynamics approach a manifold after a short
initial time. If an approximate expression for such a manifold can be found, it can be
exploited mathematically to simplify the system \citep{constable2017exploiting}.  This is
usually done by assuming that selection in the model is weak \citep{barton1995general,
nagylaki1999convergence, kirkpatrick2002general, lion2018price}. We identify the linkage
disequilibrium as a fast variable in our model, isolate it using a coordinate
transformation and find an approximation of the surface to which the dynamics converge.
Here we show that the existence of a time-scale separation
between variables and hence attraction to the QLE manifold is not 
exclusively associated with simple dynamics which are characterised by
gradient-like convergence to an interior equilibrium.

The model presented here has complex dynamics, such as
bistability and a global bifurcation. We show that, in such a system, it is
still possible to find an approximate yet accurate explicit expression for the QLE
manifold. For analytical tractability, following standard methods in
population genetics, we derive a continuous-time approximation to our discrete-time model
\citep{nagylaki1999convergence, burger2000mathematical, pontz2018evolutionary}. We use
this continuous-time approximation to find an expression for the QLE manifold. We go on to
use this to constrain the dynamics analytically to this surface, reducing the dimension of
the system.  We are then able to calculate the stability of the now-planar heteroclinic
cycle that exists in our model within certain parameter regimes.  Constraining the
dynamics is a powerful step as it allows for the use of the only known analytic
heteroclinic stability condition in discrete-time for planar cycles
\citep{hofbauer2000sophisticated}.  In the vicinity of this heteroclinic cycle, strong
fluctuations are possible on the QLE manifold.

Finally, we numerically assess the accuracy of our approximation of the QLE manifold
against both sources of error: the \textit{quasi steady-state assumption} and the use of
the continuous-time derived manifold within the discrete-time system. We find that the
manifold is a good approximation for the discrete-time system for both damped oscillations
towards the unique interior equilibrium and the approach towards the heteroclinic cycle.
}

\hl{\section{The model}}

\hl{We investigate the dynamics of haplotype frequencies of two alleles at two interacting
loci, in an infinite population, undergoing a specific selection regime (uniquely defining
the fitness matrix $W$), recombination and random union of gametes (panmixia). Once the
fitness matrix and the parameter $\delta$ are defined, the system of equations in question
is fully defined \eqref{eq:generalmodel}. First, we describe the biological processes
which justify our selection regime, then we present the resulting fitness matrix
\eqref{eq:fitnessmatrix}.

Our model describes the evolution of recombination hotspots by following the dynamics
between a modifier gene --- producing a recombinogenic protein --- and a target gene, on
which the protein binds to, causing a double-strand break and initiating recombination
\citep{ubeda2019prdm9}. This model is here re-written as a system of
selection-recombination equations. This system describes the following general processes:
a fitness benefit derived from recognition between modifier and target $(\beta)$, a
fitness cost derived from gene conversion $(\gamma)$ and the reshuffling of alleles in
double heterozygotes caused by gene conversion and crossover $(\delta)$
\citep{ubeda2019prdm9}. Our original formulation of the model included another parameter
$\alpha$, which we have normalised to one (without loss of generality) for simplicity.

The dynamics of the matching process between homozygotes and gene conversion leads to the
following system of equations describing the frequency of each haplotype in the next
generation}
\begin{equation} \label{eq:discreteTimeMainTextSystem}
    \begin{split}
        x_1' &= \frac {1}{\bar w} \biggl(x_1 [1 + \beta x_1 - \gamma x_2] - \delta D\biggr),\\
        x_2' &= \frac {1}{\bar w} \biggl(x_2 [1 - \beta x_2 + \gamma x_1] + \delta D\biggr),\\
        x_3' &= \frac {1}{\bar w} \biggl(x_3 [1 - \beta x_3 + \gamma x_4] + \delta D\biggr),\\
        x_4' &= \frac {1}{\bar w} \biggl(x_4 [1 + \beta x_4 - \gamma x_3] - \delta D\biggr),
    \end{split}
\end{equation}
where the linkage disequilibrium between alleles is
\begin{equation}
    \label{eq:linkageDisequilibrium}
    D = x_1 x_4 - x_2 x_3,
\end{equation}
and the population mean fitness is
\begin{equation}
    \label{eq:wbar}
    \bar w = x_1 + x_2 + x_3 + x_4 + \beta \left(x_1^2  - x_2^2 - x_3^2+ x_4^2\right).
\end{equation}
\hl{Superscript primes indicate the value of the variable in the next generation. The
population mean fitness, $\bar w$, ensures that the sum of the haplotype frequencies
remains constant in time. To ensure the right hand side of the difference equations does
not become negative, which would imply that the number of gametes produced is negative, we
require that the parameters $\beta$, $\gamma$ can only take values between 0 and 1. This
can be justified by the fact parameters represent probabilities in the context of the
selection-recombination equations. The parameter $\delta$ can only take values between 0
and $\tfrac{1}{2}$.

Our fitness matrix and therefore our model has similarities with that of
\citep{karlin1970linkage}. They study symmetric viability, meaning they impose a symmetric
fitness matrix. Ours is perhaps superficially similar but has a crucial difference; our
matrix is not symmetric. Our matrix results in certain local symmetries within the
resulting equations --- symmetries which are a hallmark of heteroclinic cycles. In that
sense, our model is closer to the ones of \cite{haig1991genetic} who also studied a
process with a non-symmetric fitness matrix also finding a heteroclinic cycle. We choose a
specific example to study for mathematical tractability and to link it to specific
biological examples.
}

\begin{table}
    \begin{center}
        \begin{tabular}{ c | c c } \label{tab:variableDescription}
                  & $A_1$ & $A_2$ \\ \hline
            $B_1$ & $x_1$ & $x_3$ \\
            $B_2$ & $x_2$ & $x_4$
        \end{tabular}
    \end{center}
    \caption{
            Relations between the haplotype frequencies, $x_1$, $x_2$, $x_3$, $x_4$, the
            alleles controlling the recombinogenic protein type, $A_1$, $A_2$, and the
            alleles controlling the target site sequence, $B_1$, $B_2$. The table
            indicates that the allele frequencies are obtained by summing over the
            haplotype frequencies in the corresponding row or column. Explicitly, $A_1 =
            x_1 + x_2$, $A_2 = x_3 + x_4$, $B_1 = x_1 + x_3$ and $B_2 = x_2 + x_4$.
        }
    \label{tab:Tab1}
\end{table}

\section{Analysis and Results}

\begin{figure}
    \centering
     \includegraphics[width = \textwidth]{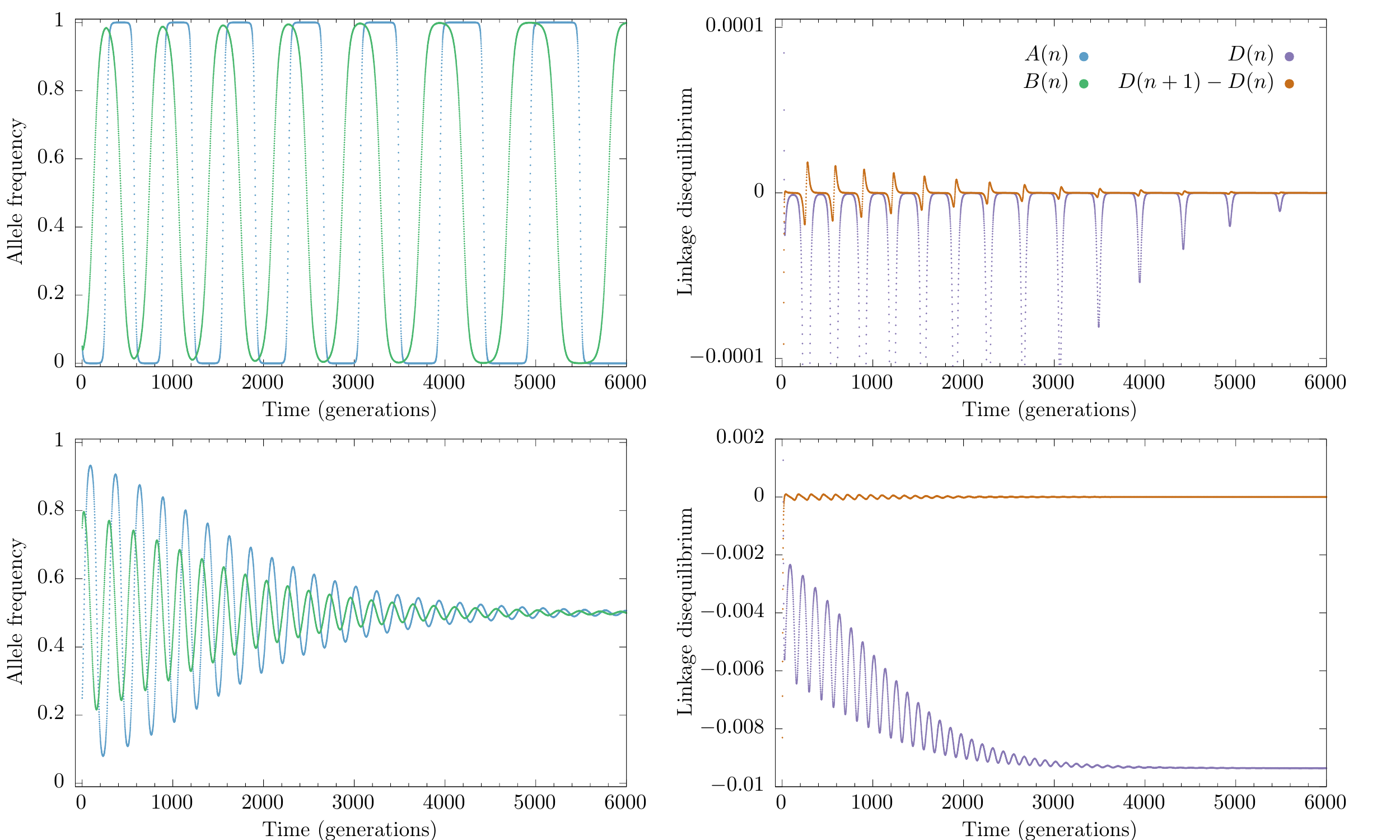}
     \caption{\textbf{Time series showing examples of the two types of behaviour of the
         discrete-time model \eqref{eq:WF-Transformed}}. The examples in the top
         row have initial conditions: $A(0) = 0.05$, $B(0) = 0.95$, $D(0) = 0.0005$ and
         those in the bottom row have initial conditions $A(0) = 0.25$, $B(0) = 0.75$,
         $D(0) = 0.0005$. Trajectories in both rows were solved with the same set of
         parameters: $\beta = 0.1$, $\gamma = 0.13$, $\delta = 0.2$. The
         top row shows a typical trajectory nearby the heteroclinic cycle. It also shows
         that after an initial period of rapid change, the linkage disequilibrium
         eventually changes relatively slowly ($D'$ becomes approximately constant in
         time), indicating the convergence of the dynamics to QLE manifold. The bottom row
         shows a typical orbit exhibiting damped oscillations and convergence to the
         asymptotically stable interior equilibrium \eqref{eq:internalEquilibrium}.
     }
\label{fig:one}
\end{figure}

The model has two different qualitative behaviours: convergence to equilibrium and
sustained oscillations. In both cases, the rate-of-change of $D$ tends towards zero on a
faster time scale than the rate-of-change of the allele frequencies (see Figure
\ref{fig:one}). This suggests that the system has two separate time scales and that the
dynamics converge towards the QLE manifold. We will find an approximate expression for
this manifold.

For brevity, we introduce $A = A_1$ and $B = B_1$ to denote the frequency of the first
recombinogenic protein and its matching target allele, respectively. The frequency of the
second recombinogenic protein and its target allele can then be written as $A_2 = 1 - A$
and $B_2 = 1 - B$.

\hl{subsection{Change of variables}}

The first step towards finding an approximation of the QLE manifold is changing
coordinates \hl{so} that they describe the allele frequencies and linkage disequilibrium. We
achieve this by transforming variables from haplotype frequencies to allele frequencies
using
\begin{equation} \label{eq:forwardTransformation}
    \begin{split}
        A&=x_1 + x_2,\\
        B&=x_1 + x_3,\\
        D&=x_1 x_4 - x_2 x_3,
    \end{split}
\end{equation}
where $A$ and $B$ take values on the interval $[0,1]$. $D$ represents linkage
disequilibrium between alleles and takes values on $[-\frac 14, \frac14 ]$.
If we consider \eqref{eq:forwardTransformation} to be the forward transformation, we
arrive at the backward transformation
\begin{equation} \label{eq:backwardTransformation}
    \begin{split}
        x_1 &= A B + D,\\
        x_2 &= A (1 - B) - D,\\
        x_3 &= (1 - A) B - D,\\
        x_4 &= (1 - A) (1 - B) + D.
    \end{split}
\end{equation}
Transforming using \eqref{eq:forwardTransformation}, the discrete-time model becomes
\begin{equation} \label{eq:WF-Transformed}
    \begin{split}
        A' &= \frac{1}{\bar{w}} \beta A (1-A)(2B- 1)+A, \\[6pt]
        B' &= \frac{1}{\bar{w}} \biggl[(\gamma - \beta)
        B (2A - 1)(B-1) + \gamma (2 B - 1) D \biggr]+B, \\[6pt]
        D' &= \frac{1}{\bar{w}^2} \biggl[ (A-1) A (B-1) B (\beta -\gamma) +\\
        &D \biggl(\beta [2A(A - 1) (B^2 - B) (\gamma + \beta) + \\
        &A(A - 1) \gamma - (2A - 1)(\delta -1) (2B - 1)] - \delta + 1\biggr) + \\
        &D^2 \biggl(\beta(\beta + \gamma)(2A - 1)(2B - 1) + \beta(-2\delta + 3) +
        \gamma\biggr) + \\
        &2 \beta  D^3 (\beta + \gamma ) \biggr].
    \end{split}
\end{equation}
Additionally, $\bar w$ is transformed into
\begin{equation}
    \bar w = 1+\beta (2A - 1)(2B - 1) + 2 \beta D.
\end{equation}

\hl{As these coordinates include linkage disequilibrium ($D$) explicitly, they allow for a
simple interpretation of the surface of total linkage equilibrium: the \textit{Wright
manifold}. This surface can now be written as the part of state space where $D = 0$
\citep{rice2004evolutionary}.}

\hl{\subsection{Equilibria and local stability}}

\hl{The system has a maximum of ten solutions when solving for potential equilibria. Five of
these live within the positive state space of the model and are therefore biologically
feasible. Four of the five biologically realistic equilibria are located
at the four vertices of the tetrahedron that forms the 3-simplex (in haplotype
coordinates). These corner equilibria, in allelic coordinates}
$(A,B,D)$, are
\begin{equation}
    \begin{split}
        \Phi_1 &= (1,1,0),\\
        \Phi_2 &= (1,0,0),\\
        \Phi_3 &= (0,1,0),\\
        \Phi_4 &= (0,0,0).
    \end{split}
\end{equation}
\hl{We analysed the linear stability of these equilibria in
\cite{ubeda2019prdm9} and we summarise the main results here. For our choice of parameters
the equilibria $\Phi_2$ and $\Phi_3$ are always unstable. Moreover, if $\beta<\gamma$
these equilibria are saddles.  The equilibria $\Phi_1$ and $\Phi_4$ are stable if
$\beta>\gamma$ and are saddles, and thus unstable, if $\beta<\gamma$. Note that if $A$ or
$B$ take values of either 0 or 1 then $D=0$. Upon inspection of the transformed models, we
find that the lines connecting the equilibria $\Phi_1$ to $\Phi_2$  $(A=1, D=0)$, $\Phi_2$
to $\Phi_4$  $(B=0, D=0)$, $\Phi_4$ to $\Phi_3$  $(A=0, D=0)$ and $\Phi_3$ to $\Phi_1$
$(B=1, D=0)$ are all invariant. When all these equilibria are saddles (i.e.\ when
$\beta<\gamma$) a heteroclinic connection exists:}
\begin{equation*}
        \dotsb \rightarrow \Phi_1 \rightarrow \Phi_2 \rightarrow \Phi_4 \rightarrow \Phi_3
        \rightarrow \Phi_1 \rightarrow \dotsb.
\end{equation*}
The fifth equilibrium is positioned in the interior of the simplex. For this interior
equilibrium it is easily verified that $\dot A = 0$ and $\dot B = 0$ for $A=B=\tfrac 12$.
The interior equilibrium, in allelic coordinates, is
\begin{equation}
    \label{eq:internalEquilibrium}
    \Phi_5 = (\tfrac 12,\tfrac 12,D^* ),
\end{equation}
where $D^*$ is the negative root of
\begin{equation}
    (\gamma - \beta) D^{*2} - \delta D^* - \tfrac{1}{16}(\gamma - \beta) = 0,
\end{equation}
given by
\begin{equation}
    D^* =\frac{\delta- \sqrt{\delta^2 + \frac{1}{4}(\gamma - \beta)^2}}
    {2(\gamma - \beta)}.
\end{equation}
The positive root is larger than $\frac{1}{4}$ for $\delta>0$ and therefore the
corresponding equilibrium has negative haplotype frequencies.

The multipliers of the discrete-time model \eqref{eq:WF-Transformed} at the
interior equilibrium $\Phi_5$ are given by
\begin{equation}
    \begin{split}
        \lambda_1 &= 1 + \frac{\gamma D^* + \sqrt{(\gamma D^*)^2 + \tfrac{1}{4}
        \beta (\beta - \gamma)}}{\bar w^*},\\
        \lambda_2 &= 1 + \frac{\gamma D^* - \sqrt{(\gamma D^*)^2 + \tfrac{1}{4}
        \beta (\beta - \gamma)}}{\bar w^*},\\
        \lambda_3 &= 1 - \frac{\delta + 2D^*(\beta - \gamma)}{\bar w^*},
    \end{split}
\end{equation}
where $\bar w^*=1+ 2 \beta D^*$ \hl{denotes the value of $\bar w$ evaluated at the interior
equilibrium \citep{ubeda2019prdm9}. The eigenvalues $\hat\lambda_i$ of the
interior equilibrium of the continuous-time approximation} are given by $\hat\lambda_i =
\lambda_i - 1$.

\hl{If $\beta>\gamma$ then $D^*>0$ and $\bar w^* >0$. Therefore, in this region of parameter
space, it is relatively easy to see that the interior equilibrium is a saddle (both in the
discrete and the continuous-time models).  Specifically, $\lambda_1$ and $\lambda_3$ are
always negative, and for $0<\delta<\tfrac 12$, $\lambda_3>-1$. $\lambda_2$ is always
positive. If $\beta<\gamma$ then $D^*<0$. Eigenvalues $\lambda_1$ and
$\lambda_2$ can now form a conjugate pair of complex eigenvalues. For the equilibrium to
be locally stable in the discrete-time model we require $|\lambda_{1,2}|<1$. This leads to
the conditions for local stability}
\begin{equation}
    \label{eq:stabilityCondition}
    2\gamma \bar w^* D^*< \frac{1}{4} \beta(\beta - \gamma).
\end{equation}
\hl{If $\delta < \tfrac {1}{2}$ this condition is always fulfilled \citep{ubeda2019prdm9}.
This stability condition \eqref{eq:stabilityCondition} applies only to the discrete-time
model as its continuous-time approximation \eqref{eq:continuousTransformedSystem} is
always locally stable (for $\beta < \gamma$).}

\hl{\subsection{Global stability: A Lyapunov function and heteroclinic cycle}}

\hl{\subsubsection{A continuous-time approximate model}}

These results on asymptotic local stability leave the question of what the global dynamics
are and, in particular, if the heteroclinic connection is an attractor, or whether orbits
\hl{move away from it. While the focus of this paper is to analyse the global stability
properties of the discrete-time model \eqref{eq:discreteTimeMainTextSystem}, we introduce
the following continuous-time approximation of the discrete-time model
\citep{nagylaki1999convergence, burger2000mathematical} to aid us in this matter
significantly
\begin{equation} \label{eq:continuousTimeMainTextSystem}
    \begin{aligned}
        \dot x_1 &= \frac{1}{\bar w} \biggl(x_1 [1 + \beta x_1 - \gamma x_2] - \delta D \biggr) - x_1, \\
        \dot x_2 &= \frac{1}{\bar w} \biggl(x_2 [1 - \beta x_2 + \gamma x_1] + \delta D \biggr) - x_2, \\
        \dot x_3 &= \frac{1}{\bar w} \biggl(x_3 [1 - \beta x_3 + \gamma x_4] + \delta D \biggr) - x_3, \\
        \dot x_4 &= \frac{1}{\bar w} \biggl(x_4 [1 + \beta x_4 - \gamma x_3] - \delta D \biggr) - x_4,
    \end{aligned}
\end{equation}
where derivatives with respect to time $t$ are denoted by a dot above a variable. The
expressions for $\bar w$ and $D$ are given by \eqref{eq:linkageDisequilibrium} and
\eqref{eq:wbar}, the same as in the discrete-time model. The continuous-time model written
in the transformed variables is
\begin{equation}
    \begin{split} \label{eq:continuousTransformedSystem}
        \dot A &= \frac{1}{\bar{w}} \beta A (1-A)(2B - 1), \\[6pt]
        \dot B &= \frac{1}{\bar{w}} \biggl[(\gamma - \beta)
        B (2A - 1)(B-1) + \gamma (2 B - 1) D \biggr], \\[6pt]
        \dot D &= \frac{1}{\bar{w}} \biggl[ (\gamma-\beta)
        \left[D^2-A B (1-A)(1-B) \right]-
        \beta D (2 A - 1)(2 B - 1) - \delta D \biggr].
    \end{split}
\end{equation}
It is easy to show that the equilibria for the discrete-time model and its continuous-time
approximation are the same \citep{burger2000mathematical}.  Similarly, it is easy to show
that the eigenvalues of the Jacobian at each equilibrium in the continuous-time model
equal the discrete-time eigenvalues minus unity --- a consequence of the fixed time-step
in the discrete-time system. We use the continuous-time model in two ways: introducing a
Lyapunov function for the interior equilibrium, showing it to be globally stable; using it
to find an analytically tractable version of the approximate QLE manifold, as the
expression is significantly simpler when derived from the continuous-time model.

\subsubsection{Lyapunov function}

For the continuous-time model it is relatively easy to show that the heteroclinic cycle
repels orbits using a Lyapunov function. Before we show this, we first observe that for
any solution of (\ref{eq:continuousTransformedSystem}) as long as $D \le 0$ at some point
in time, $D \le 0$ onwards if $\beta<\gamma$, and with equality only if the solution lives
on the heteroclinic connection. This can easily be seen by inspecting the right hand side
of the differential equation describing the change in $D$ when $\beta < \gamma$, which is
negative everywhere on the Wright manifold, apart from on the heteroclinic connection,
where it is zero. Therefore, if $D(t_0) < 0$, then $D(t) < 0$ for all $t > t_0$. This
means that trajectories can pass through the Wright manifold where $D = 0$ in only one
direction, and are then confined to the region where $D \le0$ once they have
done so.}

\begin{figure}
    \centering
   \includegraphics[width=\textwidth]{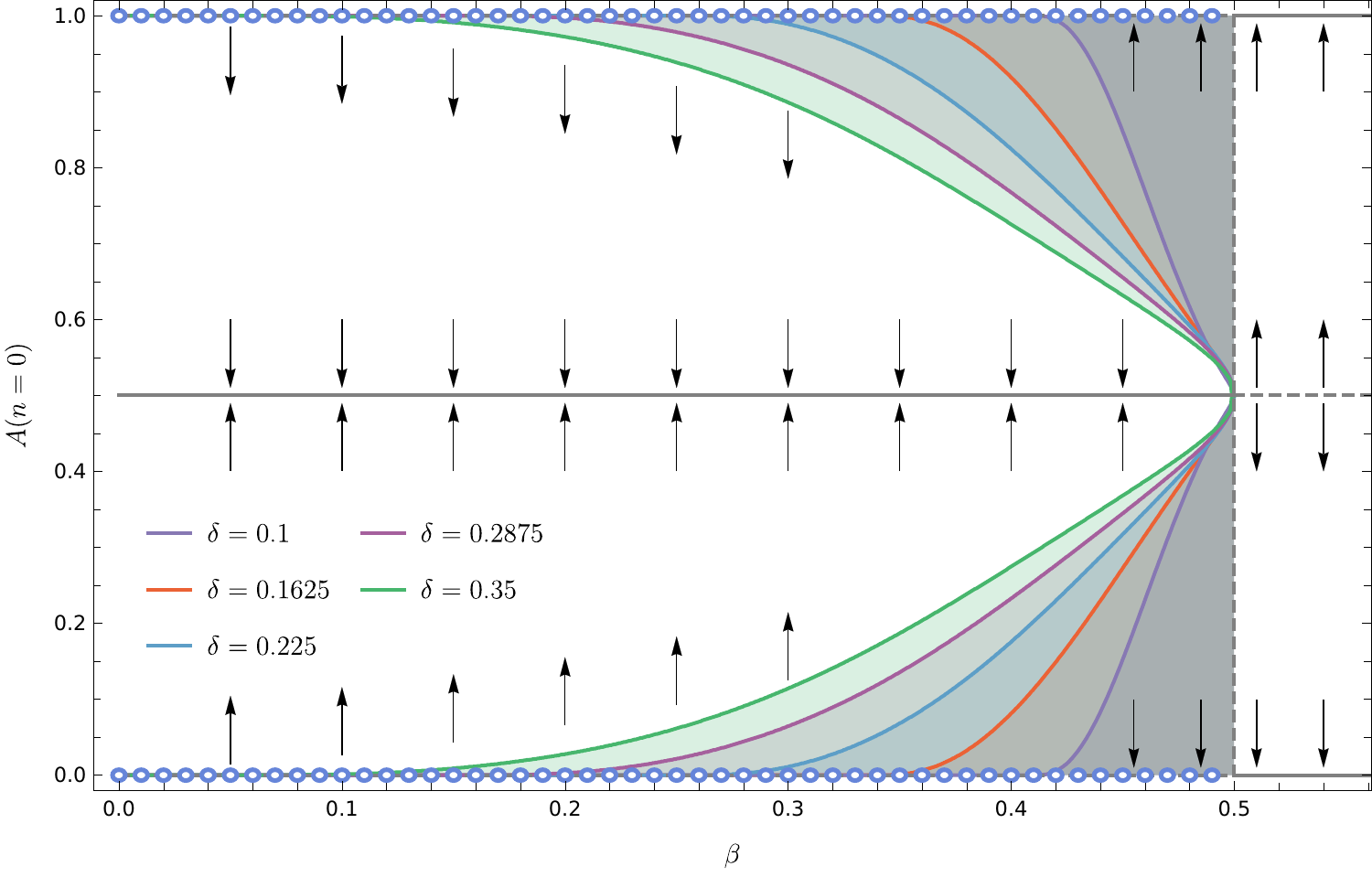}
   \caption{\textbf{The basin of attraction of the heteroclinic cycle
        against $\beta$ for the discrete-time model}.
        The diagram shows the different qualitative behaviours of the model resulting from
        different initial conditions. The arrows point towards the different attractors.
        The shaded regions show the basins of attraction of heteroclinic cycle for varying
        values of $\delta$ (see legend).  The diagram was constructed by starting orbits
        at different initial conditions, sampled at equally spaced intervals along the
        line connecting the equilibria $\Phi_1$ and $\Phi_4$ for which $A = B$ and $D =
        A(1-A)$ in allelic coordinates, or $(x_1, 0, 0, 1 - x_1)$ in gametic coordinates.
        We determine whether a specific orbit reaches interior equilibrium or a
        heteroclinic cycle numerically: if an orbit reaches within $\epsilon = 10^{-12}$
        distance from the equilibrium, it is assumed to be at equilibrium.  The first
        trajectory moving along the line of initial conditions which does not tend towards
        equilibrium is taken to be on the basin boundary. The heteroclinic cycle exists on
        the left of the vertical dashed line at $\beta = \gamma = 0.5$. At this point both the
        interior equilibrium and heteroclinic cycle lose stability and all trajectories
        tend toward one of the corner equilibria, $\Phi_1$ or $\Phi_4$. Parameters:
        $\gamma=0.5$, $\delta$ as indicated in figure. Dashed lines represent unstable
        equilibria, drawn lines represent stable equilibria and small blue circles represent
        the heteroclinic cycles.}
    \label{fig:original_basin}
\end{figure}

With this established, we now consider the function
\begin{equation}
    \label{eq:lyapunovFunction}
    V(A,B) = [A(1-A)]^{\gamma - \beta} [B(1-B)]^{\beta}.
\end{equation}
\hl{This function \eqref{eq:lyapunovFunction} serves as a natural candidate for a Lyapunov
function of system \eqref{eq:continuousTimeMainTextSystem} as it retains invariance of the
system along the boundaries (where either $A = 0$, $A=1$, $B = 0$ or $B=1$). Indeed, for
$\beta<\gamma$ this function takes the value $V=0$ along the heteroclinic connection, and
takes positive values anywhere else in or on the simplex. The continuous-time model with
$D$ set to zero \eqref{eq:continuousTransformedSystem} is equivalent to the replicator
equations for $2 \times 2$ games and our Lyapunov function \eqref{eq:lyapunovFunction} is
equivalent to that of this system, serving as its constant of motion
\citep{hofbauer1998evolutionary}.

The candidate function $V$ is a Lyapunov function if $\beta <\gamma$ for
orbits which at some point pass through the Wright manifold. To show this, we inspect its
time derivative along solutions of}
(\ref{eq:continuousTransformedSystem}):
\begin{equation} \label{eq:lyapunovDerivative}
    \dot V = - \beta \gamma\frac{D}{\bar w}  \frac{(1-2B)^2}{B(1 - B)}V.
\end{equation}
\hl{The right hand side of (\ref{eq:lyapunovDerivative}) is always less than or equal to zero
if $D\le 0$, meaning $V$ is a Lyapunov function within this region. For orbits starting in
the forward invariant part of state space where $D < 0$ the value of $V$ will thus
increase or stay constant over time. The $\omega$-limit of these orbits must therefore be
invariant sets for which either $D=0$ or $B= \tfrac{1}{2}$. If $\beta<\gamma$ the only
invariant part of the Wright manifold $D=0$ is the heteroclinic connection, where $V=0$.
As the value of $V$ cannot decrease and is positive for all points in or on the simplex
that are not part of the heteroclinic connection, the heteroclinic connection cannot be an
$\omega$-limit of these orbits, within which the only other candidates are the invariant
sets contained with $B=\tfrac 12$, which is the interior equilibrium $\Phi_5$. Any orbits
starting within the parts of the simplex where $D<0$ will therefore move towards the
interior equilibrium.

A corollary of this observation is that arbitrarily close to the heteroclinic connection,
where $D=0$, there will be points that are within the region of the simplex where $D<0$.
The Lyapunov function \eqref{eq:lyapunovFunction} shows that orbits starting at these
points will move away from the heteroclinic connection, towards the interior equilibrium.
The heteroclinic connection is therefore not stable. The interior equilibrium clearly is
stable and must be the attractor for all initial points in the interior of the simplex for
which initially $D<0$.  This shows that in the continuous-time model the heteroclinic
cycle is unstable.  Simulations suggest that the interior equilibrium is a global
attractor within the simplex.

\subsubsection{Discrete-time heteroclinic cycle}

The Lyapunov argument does not carry over to the discrete-time model. In the discrete-time
model, does the heteroclinic connection attract or repel? We analytically investigate this
using the approximate QLE manifold in section \ref{ch:HCStability}. We also numerically
investigate the regions of initial condition space in which the cycle is attracting, and
the results are plotted in Figure \ref{fig:original_basin}. In the diagram we can
distinguish two regions in parameter space with qualitatively different behaviour, and the
boundary between them:}
\begin{enumerate}
    \item Within the first region, $\beta < \gamma$, the interior equilibrium is stable
        and attracts nearby orbits. Within this region the heteroclinic connection
        also attracts. Between the two attractors we find the boundary of the basins of
        attraction. The basin boundary moves towards the heteroclinic connection for small
        $\beta$.

    \item Within the second region $\beta > \gamma$. All trajectories converge to one
        of the corner equilibria, $\Phi_1$ or $\Phi_4$, apart from orbits starting
        exactly at the unstable interior equilibrium $\Phi_5.$

    \item Between these two regions $\beta = \gamma$, all trajectories converge to the
        \hl{Wright manifold. On the Wright manifold there is a line of unstable
        equilibria for which $B=\tfrac 12$, $D=0$. Orbits starting on the
        Wright manifold with $B<\tfrac 12$ converge to the line $A=0$, $D=0$, and those
        starting with $B>\tfrac 12$ converge to the line $A=1$, $D=0.$}
\end{enumerate}
These results show that the heteroclinic connection in the discrete-time model can be
stable. To find out how general this is we will next analytically determine the stability
\hl{of the heteroclinic connection in the discrete-time model. First, we
approximate the QLE manifold towards which the trajectories converge.}

\subsection{The QLE manifold}

\begin{figure}
    \centering
    \includegraphics[width = \textwidth]{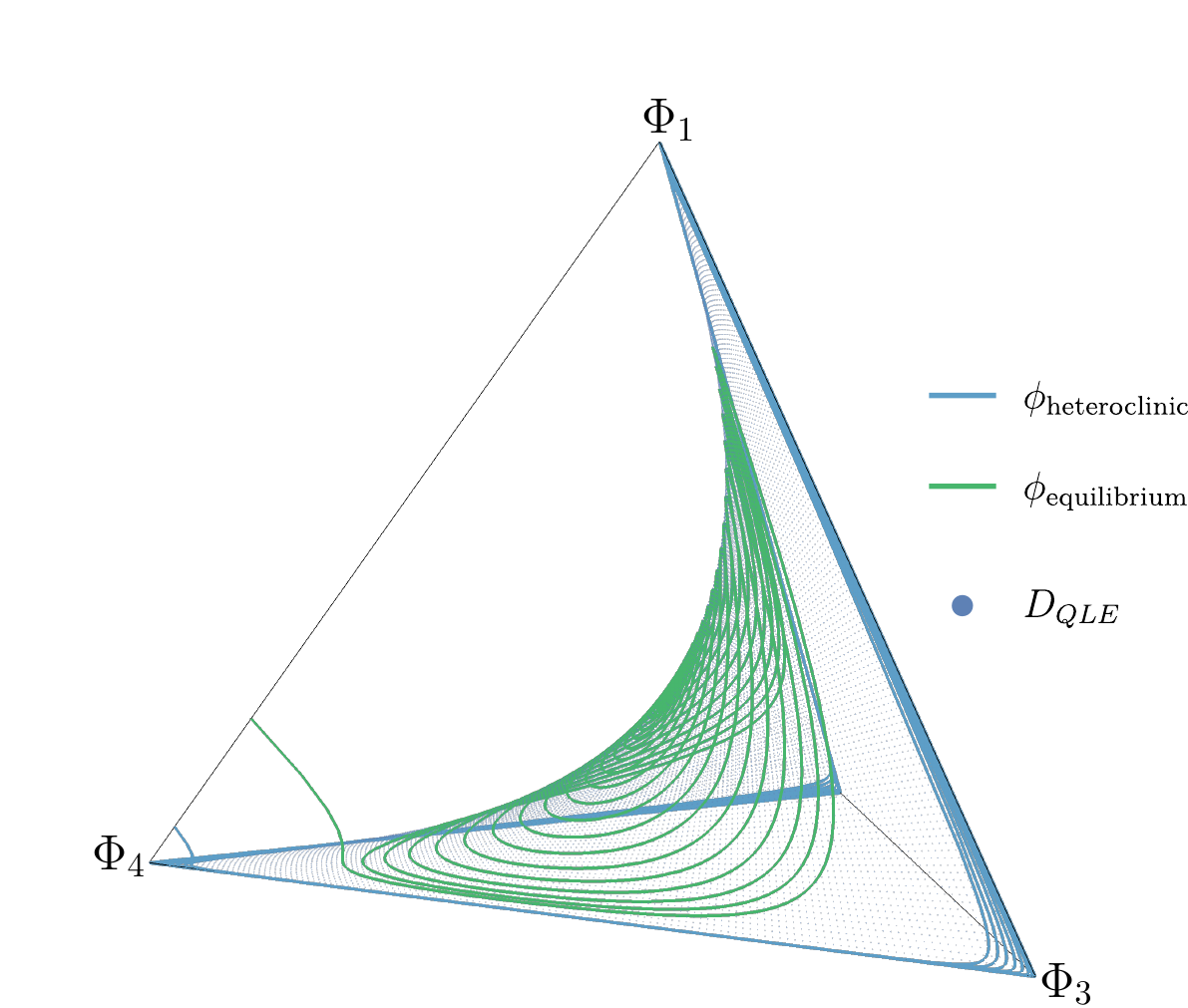}
    \caption{\textbf{The approximate quasi-linkage equilibrium manifold, and the approach
        to it by two typical trajectories of the discrete-time model.} Two trajectories,
        $\phi_{\text{heteroclinic}}$ and $\phi_{\text{equilibrium}}$, differing only in
        initial conditions, of the transformed discrete-time
        \eqref{eq:discreteTimeMainTextSystem} system within the tetrahedron, both
        converging quickly to a slow manifold. Here, the small dots are
        points on the manifold $D_{QLE}$, given by \eqref{eq:QLEManifold}. As can be
        seen, the trajectories converge quickly to this manifold. Parameters and initial
        conditions as in Figure \ref{fig:one}.}
     \label{fig:manifold_two_trajectories}
\end{figure}

If $\beta = \gamma$ the interior equilibrium is degenerate: in the discrete-time model the
equilibrium has two real multipliers at unity (whilst the interior equilibrium of the
continuous-time model has two eigenvalues at zero). Because there are two eigenvalues at
unity (zero), the equilibrium will have a two dimensional center manifold. If $\beta =
\gamma$ the center manifold is the Wright manifold, the part of state
space where $D = 0$, and where the gamete frequencies are in linkage
equilibrium. The third eigenvalue has a modulus smaller than one (smaller than zero for
the continuous-time model) and the associated stable manifold is given by the line $A = B
= \frac{1}{2}$. Orbits on this stable manifold move towards the center
manifold.

If $\beta < \gamma$ these two multipliers become a complex pair with real part smaller
\hl{than one (or negative real part for the continuous-time model). The equilibrium within
this region is hyperbolic (for all $0 < \delta < \frac{1}{2}$) for the ODE
\eqref{eq:continuousTransformedSystem}. The same is true for the map
\eqref{eq:WF-Transformed} when there is not equality in the stability condition
\eqref{eq:stabilityCondition}. The center manifold morphs into a two
dimensional invariant manifold that is different from the Wright manifold and contains the
interior equilibrium \eqref{eq:internalEquilibrium}. On this manifold, orbits cycle around
the equilibrium. The invariant manifold containing the third eigenvector, the line on
which $A = B = \tfrac{1}{2}$, remains in existence. Over this line, orbits quickly
converge towards the equilibrium and as they approach the linkage disequilibrium, $D$
changes rapidly while the allele frequencies $A$ and $B$ remain unchanged. Other orbits
show a similar behaviour (see Figure \ref{fig:manifold_two_trajectories}): orbits
generally converge towards the two dimensional manifold. Once orbits are close to this
manifold the orbits move slowly towards either the interior equilibrium or the
heteroclinic cycle, depending on the initial conditions (see 
Figure~\ref{fig:original_basin}).}

\begin{figure}
    \centering
    \includegraphics[width=\textwidth]{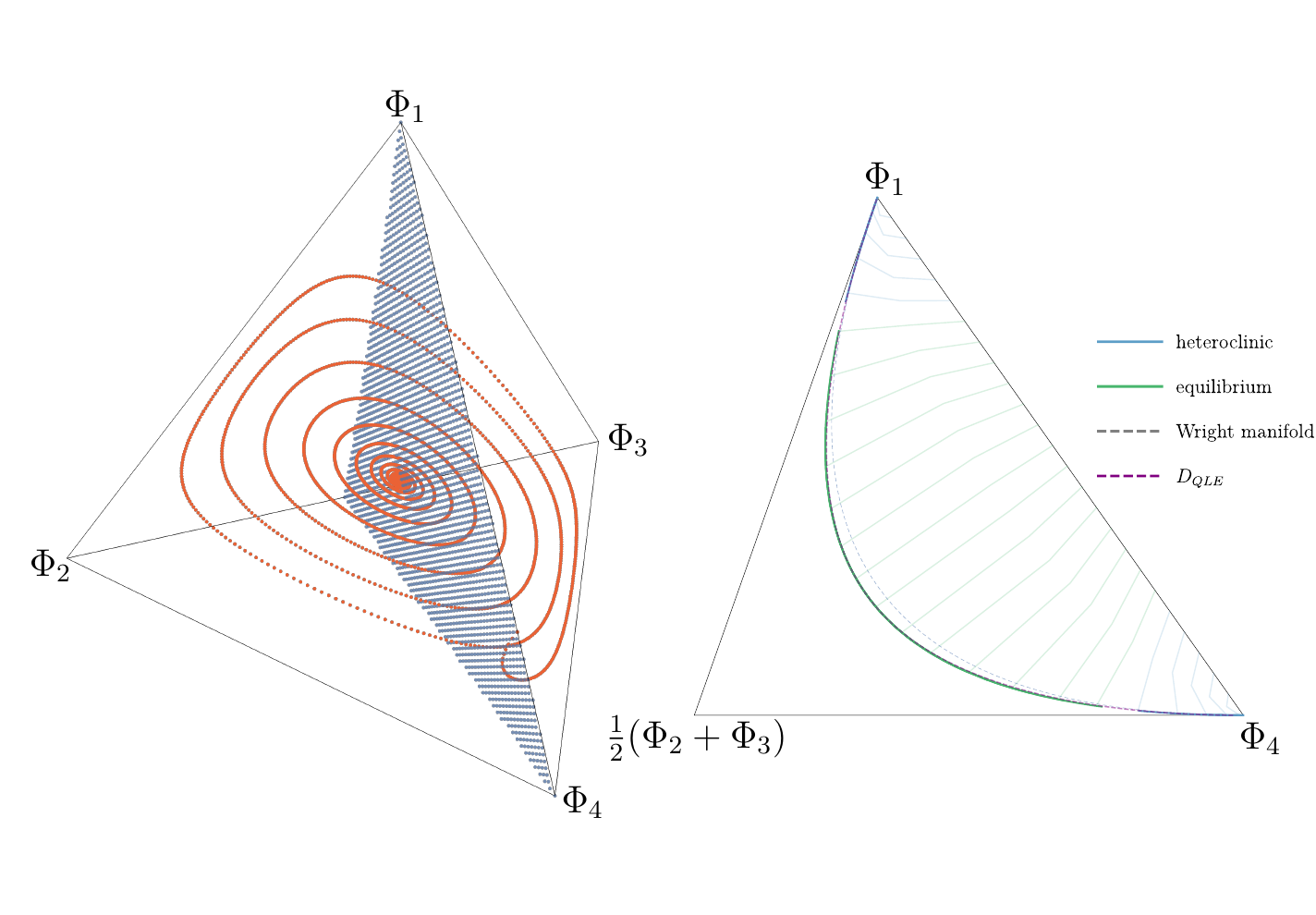}
    \caption{\hl{\textbf{The fast approach to the QLE manifold shown using a Poincaré
        section}.  The dynamics of our model has two different times scales and shows
        slow-fast dynamics. (a) A typical trajectory of the model}
        \eqref{eq:discreteTimeMainTextSystem}, simulated using $\beta = 0.1, \gamma =
        0.13$ and $\delta = 0.11$ and initial conditions $(x_1(0), x_2(0), x_3(0), x_4(0))
        = (0.24, 0, 0, 0.76)$. To visualise the slow-fast dynamics we following the
        Poincar\'e section $x_2=x_3$ (=$A=B$) and record every instance where the orbit
        (shown in red) cuts through this section. (b) The intersection points for a orbit
        plotted on the Poincar\'e section. The points of intersection of 22 trajectories
        are shown. The trajectories have initial conditions equally spaced on the line
        connecting $\Phi_1$ to $\Phi_4$. The parameters used are $\beta = 0.3, \gamma =
        0.35$ and $ \delta = 0.2$. The figure shows the fast approach towards the slow
        manifold (the thin, drawn lines connect the points of intersection from the same
        initial condition). The slow manifold is visible as the accumulation of points
        \hl{forming a curve.  Although the true slow manifold (blue and green filled lines) and
        our approximation, DQLE, (purple dashed line) are distinct from the Wright
        manifold (dashed grey line) apart from at the corners, where they intersect, they
        are very close and the purple curve is covered by the blue and green line in most
        of the figure.  Green dots are from orbits that end up in the interior
        equilibrium, $\Phi_5$, blue dots from orbits going towards the heteroclinic cycle.
        The gap on the slow manifold between the blue and green points contains the basin
        boundary. There will be an invariant closed curve located on the slow manifold in the
        middle of this gap.}}
    \label{fig:poincareSection}
\end{figure}

\hl{To approximate the QLE manifold, we will use a quasi-steady state argument. Specifically,
we say that the change in linkage disequilibrium $D(t)$ occurs on a much faster time scale
than changes in the allele frequencies and will therefore settle on a quasi-equilibrium.
This means that we can assume that the allele frequencies $A$ and $B$ are effectively
constant, as $D$ settles. With this assumption, we then solve the equilibrium equation for
$D$ as a function of the allele frequencies, $D_{QLE}(A,B)$. It turns out that this gives
a good approximation for the QLE manifold for the discrete-time model as well as the
continuous-time approximation.

Simulations suggest that the gamete frequencies are attracted towards the manifold where
they are in quasi-linkage equilibrium. We approximate the QLE manifold by}
\begin{equation} \label{eq:QLEManifold}
    \begin{split}
        D_{QLE}(A,B) &= \frac{\beta (2 A - 1)(2 B - 1) + \delta }
        {2(\gamma-\beta)}\\[6pt]
        &- \sqrt{\left(\frac{\beta (2 A - 1)(2 B - 1) + \delta}
        {2(\gamma-\beta)}\right)^2+A B (1-A)(1-B)}.
    \end{split}
\end{equation}
\hl{As we show in \ref{sc:AppendixB} the relevant slow time-scale is proportional to $(\gamma
- \beta)^{-\frac{1}{2}}$.}

\subsection{Simplification by reducing to allele frequencies}

\hl{Given the tendency of the haplotype frequencies to settle on the QLE, one would expect}
that if $\gamma > \beta$, the dynamics proceed to the QLE manifold, and that the allele
frequencies then change slowly, either towards, or away from the interior equilibrium.
This is indeed what happens in the vicinity of the interior equilibrium. Further away from
equilibrium, and in particular in the vicinity of the heteroclinic cycle, this is not
\hl{necessarily true. It is possible that the manifold $D=D_{QLE}(A,B)$ is situated outside
the simplex in which all gamete frequencies are positive. If that is the case, the
dynamics will be constrained by the edges of the simplex.

Inside the simplex, $D_{QLE} \le 0$ if $\gamma > \beta$. If the manifold, $D_{QLE}$, cuts
through the sides of the simplex, it can only be on the faces where $D \le 0$, which is
when $x_1 \le 0$ or $x_4 \le 0$. In terms of allele frequencies $(A, B, D)$, that is when
$D = -AB$ or when $D=-(1-A)(1-B).$ The approximate manifold to which the dynamics are
drawn is thus given by $D=D_S(A,B)$, where}
\begin{equation}
    D_S(A,B)=\max\biggl[D_{QLE}(A,B),-AB, -(1-A)(1-B)\biggr],
\end{equation}
and we will use this to simplify the dynamics; in particular we will use it to determine
the stability of the heteroclinic cycle.

\begin{figure}
    \centering
    \includegraphics[width = \textwidth]{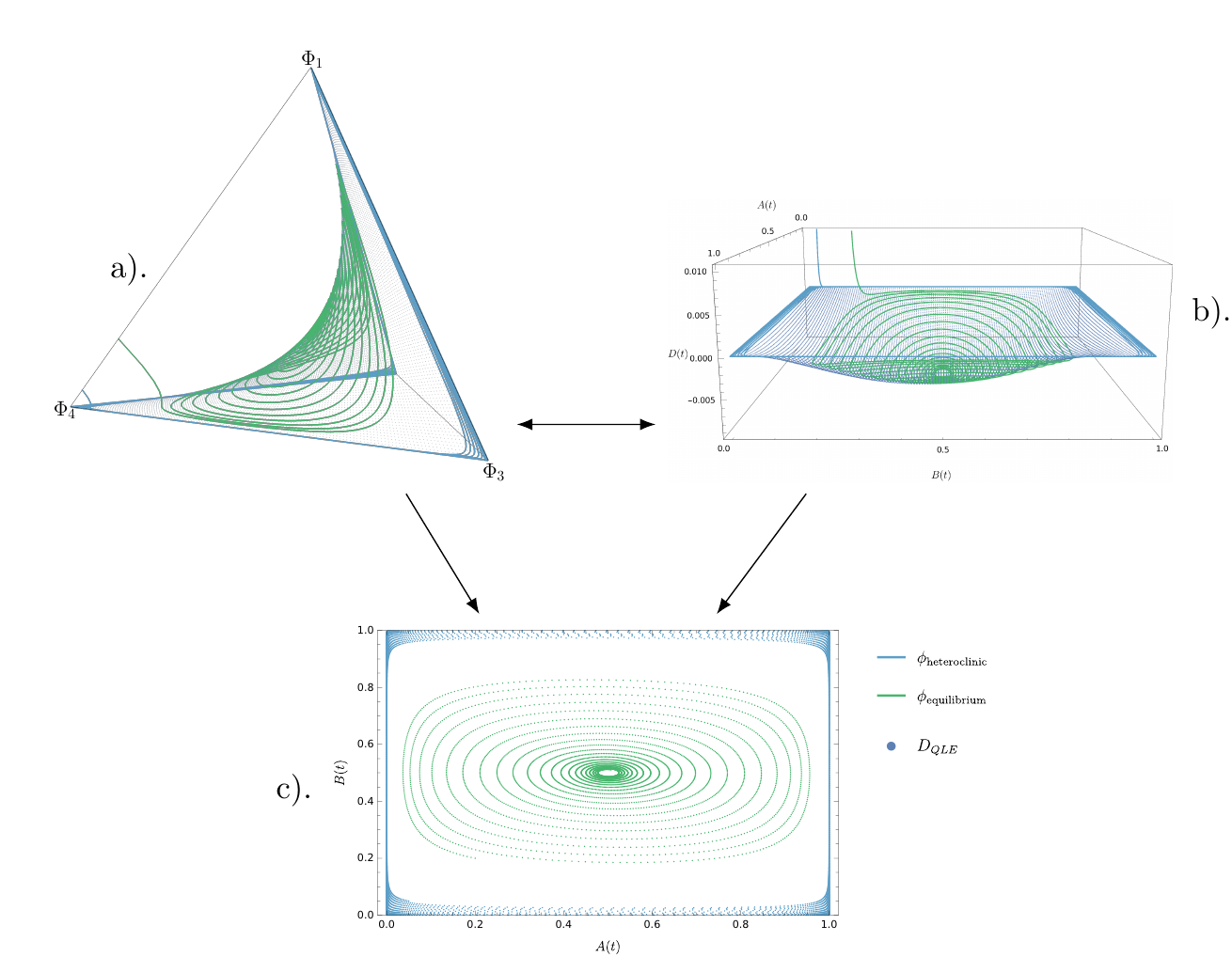}
    \caption{\textbf{The simplification of the system by using the \hl{approximate} slow
        manifold, $\mathbf{D_{QLE}}$.}
        (a) The trajectories of our model represented gamete frequencies as given by eqns
        (\ref{eq:discreteTimeMainTextSystem}), plotted on the 3-simplex. The QLE
        manifold, $D=D_{QLE}$, is also plotted with a grid of equally spaced points. (b)
        The same trajectories and the attracting manifold  plotted for the transformed
        model (\ref{eq:WFTransformedSystem}); in both panels (a) and (b) the fast approach
        to the slow manifold is visible. (c) The same trajectories but  plotted on the
        QLE manifold. The system is  reduced to a planar system in the allele
        coordinates. Parameters and initial conditions as in Figures~\ref{fig:one} and
        \ref{fig:manifold_two_trajectories}. Panel (a) is a re-use of
        Figure~\ref{fig:manifold_two_trajectories}.}
\label{fig:five}
\end{figure}

The system constrained to the attracting manifold is given by just two equations,
describing the frequencies of $A$ and $B$ on the slow manifold,
\begin{equation}
    \begin{split} \label{eq:WFTransformedSystem}
        A' &= \frac{1}{\bar{w}} \beta A (1-A)(2B - 1)+A,\\[6pt]
        B' &= \frac{1}{\bar{w}} \biggl[(\gamma - \beta)
        B (2A - 1)(B-1) + \gamma (2 B - 1) D_{S}(A,B) \biggr]+B,
    \end{split}
\end{equation}
where
\begin{equation}
    \bar{w} = \beta (2A - 1)(2B -1) + 2\beta D_{S}(A,B) + 1.
\end{equation}
The dimensionality is now reduced and the system is significantly simplified.  We can now
study and depict our model as a two dimensional system (Figure \ref{fig:five}).  The
stability of the heteroclinic cycle is governed by the magnitude of the eigenvalues in the
connected saddles that make up the cycle. In the planar system this is relatively simple
to do.

\subsection{Stability of heteroclinic cycle in the discrete-time model}
\label{ch:HCStability}

To study the stability of our heteroclinic cycle, we use the condition derived in
\cite{hofbauer2000sophisticated} which determines whether a planar discrete-time
heteroclinic cycle is attracting or not. The condition involves the product of the ratio
of the logarithm of the expanding ($e_i$) eigenvalues and the absolute value of the
logarithm of the contracting eigenvalues ($c_i$) at the saddle equilibria ($\Phi_i$ where
$i = 1,...,4$) the heteroclinic cycle travels between. We follow their notation and use
$\rho_{i}$ to denote each individual ratio and $\rho$ to denote the product of the
$\rho_{i}$,
\begin{equation}
    \begin{split}
        \centering
        \rho &= \prod_{i=1}^n \rho_{i}, \\[6pt]
        \rho_{i} &= \frac{\log e_i}
        {\lvert \log c_i \rvert}, \qquad
        i = 1,...,n.
    \end{split}
\end{equation}
\hl{For our model, $n = 4$ and therefore $\rho = \rho_1 \rho_2 \rho_3 \rho_4$. We are then
able to state the stability condition: a planar discrete-time heteroclinic cycle is
asymptotically stable if $\rho < 1$ and is unstable if $\rho > 1$
\citep{hofbauer2000sophisticated}. The specific eigenvalues for the equilibria and their
type are given in Table \ref{tab:discreteSystemEigenvalues}.  Their derivation can be
found in \ref{sc:AppendixC}.}
\begin{table}[h]
    \centering
    \begin{tabular}{c | c | c | c | c }
        Eigenvalue & $\frac{1}{1 + \beta}$ & $\frac{1 + \gamma}{1 + \beta}$ & $\frac{1}{1 - \beta}$ &
        $\frac{1 - \gamma}{1 - \beta}$  \\[6pt] \hline
        Type & $c_{1}, c_{4}$ & $e_{1}, e_{4}$ & $ e_{2}, e_{3}$ & $ c_{2}, c_{3}$  \\[6pt] \hline
        Equilibria & \multicolumn{2}{c}{$\Phi_{1}$ \& $\Phi_{4}$} \vline &
        \multicolumn{2}{c}{$\Phi_{2}$ \& $\Phi_{3}$}
    \end{tabular}
    \caption{The eigenvalues of the saddle equilibria between which the heteroclinic cycle
        travels, used to determine the asymptotic stability of the heteroclinic cycle in
        discrete-time. Eigenvalues of type $c$ are contracting (incoming), ones of type
        $e$ are expanding (outgoing).  Due to the symmetries in our system, the
        eigenvalues at $\Phi_1$ and at $\Phi_4$ are equal and the eigenvalues at $\Phi_2$ and
        at $\Phi_3$ are equal.}
       \label{tab:discreteSystemEigenvalues}
\end{table}

Calculating $\rho$ using the eigenvalues in Table \ref{tab:discreteSystemEigenvalues}, we
arrive at the condition for stability of the heteroclinic cycle
\begin{equation} \label{eq:HCstabilitydiscrete}
    \left(\frac{\log \frac{1 + \gamma}{1 + \beta}}{\lvert \log
    \frac{1}{1 + \beta}\rvert}\frac{\log \frac{1}{1 - \beta}}
    {\lvert\log \frac{1 - \gamma}{1 - \beta}\rvert}\right)^2<1,
\end{equation}
which, if $\beta < \gamma$, can be rewritten as
\begin{equation} \label{eq:HCstabilitydiscreteRewritten}
    \frac{\log (1 + \beta)}{ \log (1 - \beta)}<\frac{\log (1 + \gamma)}
    {\log (1 - \gamma)}.
\end{equation}
\hl{In this form, it is readily seen that \eqref{eq:HCstabilitydiscrete} is always satisfied
if $\beta < \gamma.$ Therefore, in our discrete-time model constrained to the QLE manifold
\eqref{eq:WFTransformedSystem}, the heteroclinic cycle is always asymptotically stable if
it exists.}

\begin{figure}
    \centering
    \includegraphics[width = \textwidth]{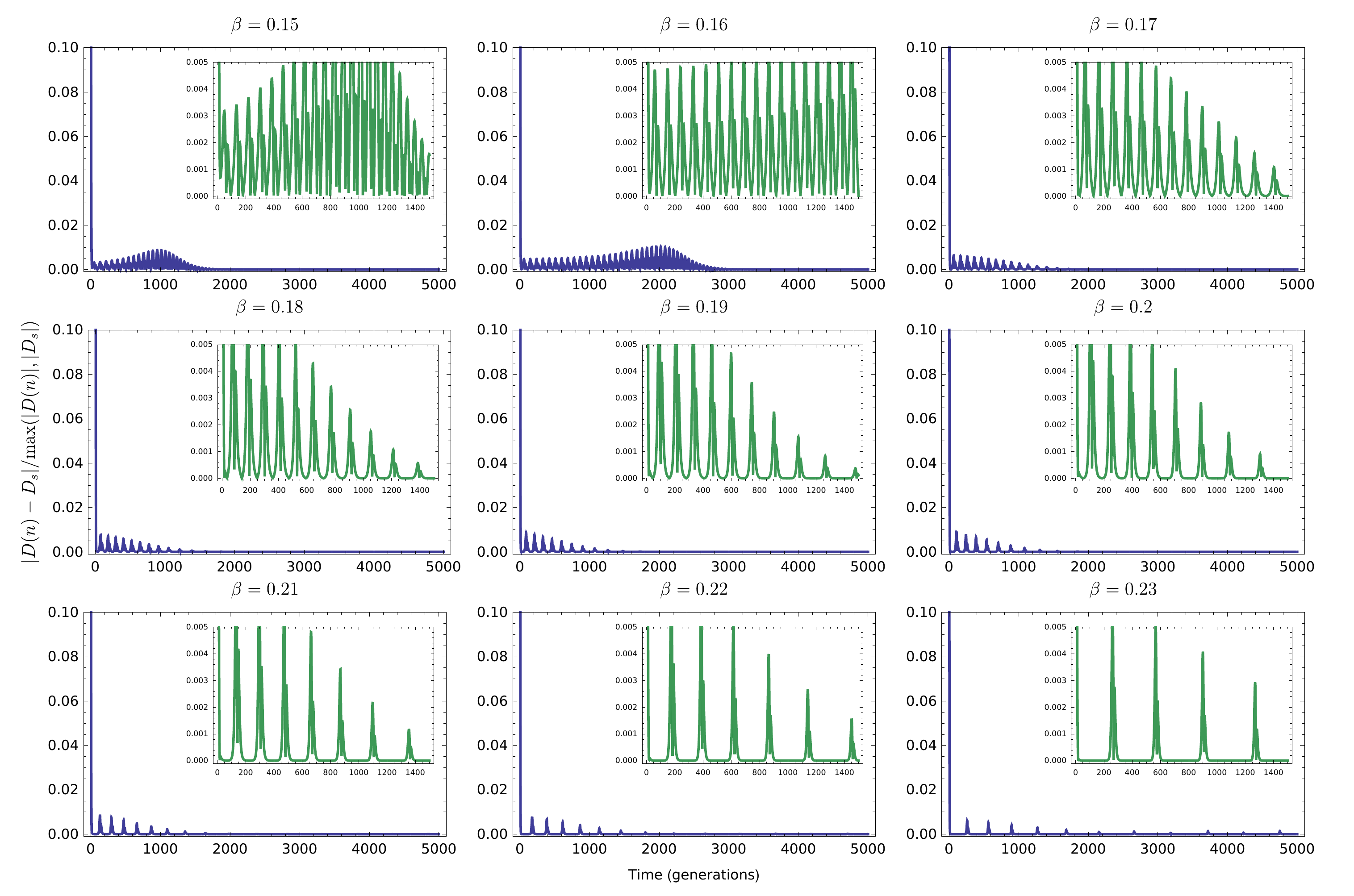}
    \caption{\textbf{Relative error of our approximate manifold $D_S$.} To justify the use
        of the manifold derived from the continuous-time system, $D_S$, we numerically
        compute the relative error between the manifold and the $D$ component of an orbit
        of the discrete time system close to heteroclinic cycle. We compute both the
        manifold expression and the orbit at the generation times of the discrete-time
        model, $n$ and plot the following error expressions $|D(n) - D_s|
        /\max{(|D(n),|D_s|)}$. Parameters were set to: $\gamma = 0.25$, $\delta = 0.3$,
        $A(0) = 0.9$, $B(0) = 0.9$, $D(0) = 0.05$ and the values of $\beta$ are indicated
        in the plot titles. The The insets show the same curves but with finer grain
        x-axis and y-axis scales allowing the bursts to be seen in more detail. The
        magnitude of error is always very low.}
\label{fig:six}
\end{figure}

\subsection{Justifying the use of $D_{S}$ derived from the continuous-time model}

\hl{In Figure~\ref{fig:six}, we show the relative error between the value of $D(n)$, the
linkage disequilibrium within the discrete-time model
\eqref{eq:discreteTimeMainTextSystem}, and $D_S(t)$, the approximate slow manifold derived
using the continuous-time approximation of the discrete-time system, finding the
difference to be small. The error is computed using
\begin{equation}
    E = \frac{|D(n) - D_S|}{\max{(|D(n),|D_S|)}},
\end{equation}
a modified form of the relative error between the approximate manifold $D_S$, and the
$D$-component of a trajectory of the discrete-time system, which aims to avoid division by
zero when one of the quantities is very small.  The standard relative error expression
could be problematic in this case, since the orbits are close to the manifold.  We produce
a time series of the distance between the $D$-component of the discrete-time orbit and the
value of $D_S$ evaluated at the values of the other variables along the orbit.  This
indicates that the continuous-time manifold, $D_S$, provides a good approximation for the
discrete-time dynamics.}

\section{Discussion}
\label{sec:discussion}

\hl{We studied a genetic system with viability selection and gene conversion that encompass a
wide range of variants where selection can be derived from different aspects of the
recombinational process \citep{ubeda2011red, ubeda2019prdm9}. We show that the selection
regime associated with a fitness benefit derived from a sequence recognition ($\beta$), a
fitness cost derived from a gene conversion ($\gamma$) altogether with the reshuffling of
alleles in double heterozygotes induced by gene conversion and crossover ($\delta$), can
lead to stable cycling dynamics in the two-locus, two-alleles model. Our model is most
similar to that of \cite{haig1991genetic}, because in both models the often assumed
symmetry of the fitness matrix \citep{karlin1970linkage} is broken. The fluctuations that
feature in the model are caused by selection for one allele burning out a target sequence
followed by selection for an alternative allele that can burn out the sequence that
replaced the old one. This pattern can repeat indefinitely and the resulting dynamics form
a heteroclinic cycle \citep{ubeda2019prdm9}. To find out if sustained fluctuations are
possible in either of our model variants we investigated whether the heteroclinic cycle
attracts or repels \citep{hofbauer2000sophisticated}.

We found that haplotype frequencies settle quickly on a state depending on the allele
frequencies in the population, and the allele frequencies change on a slower time scale
than the linkage disequilibrium \citep{kimura1965attainment}. After identifying the
linkage disequilibrium $D$ as a good candidate for the fast variable, we performed the
nonlinear change of variables from haplotype to allele frequencies, which introduces
$D(t)$ as an explicit variable. We then apply a quasi-steady state assumption to $D(t)$
and solve the resulting algebraic equation for $D$, which we use to reduce the dimension
of our system by removing dependency on $D$ altogether (Figure~\ref{fig:five})
\citep{kuehn2015multiple}. We find that the dynamics don't necessarily converge to a
single stable interior (polymorphic) equilibrium. We thus provide a biological example of
a doubly degenerate system that admits cycling.

After reducing the dimensionality, we found explicit conditions for stability of the
heteroclinic cycles. Namely, the discrete-time model allows a heteroclinic
cycle that is stable if $\beta < \gamma$; on the other hand, its continuous-time
approximation has a heteroclinic cycle that is always unstable and the
dynamics eventually settle on an equilibrium. Furthermore, we established numerically the
basin of attraction for the heteroclinic cycle and studied the accuracy of the closed-form
approximation $D_S$ of the QLE manifold used to constrain the dynamics (Figure
\ref{fig:six}).

The equilibria of the discrete and continuous-time models are the same
\citep{burger2000mathematical}.  However, the stability of the heteroclinic cycle differs
between the two models: the discrete-time model can have an attracting heteroclinic cycle
and a stable equilibrium, and thus has a region of bistability in parameter space;
however, its continuous-time approximation has, in the same region of parameter space,
$\beta < \gamma$, a globally attracting interior equilibrium point. From a dynamical
systems point of view this is not a surprise: it is well known that similar nonlinear
discrete and continuous-time models can differ in various ways \citep{may1976simple}.

However, preliminary results show that if the population in the model is finite and
multinomial sampling is used to pick the individuals who mate and are replaced
\citep{wright1969evolution,ubeda2019prdm9} --- producing a stochastic and more
biologically realistic version of our model --- we see the gap between the discrete-time
model and continuous-time approximation bridged.  Indeed, similar oscillatory behaviour is
now observed in both models. In fact, we see the two models behaving almost identically
when the population is finite, just differing in time scale. We also observe that the
deterministic slow manifold, $D_{QLE}$, is a good approximation for the dynamics of the
stochastic model, as shown to be possible in some systems by
\citep{constable2017exploiting}. An in depth analysis of the stochastic model however, is
beyond the scope of this paper. Further work could use $D_{QLE}$ to simplify the dynamics
of the stochastic implementation of the model. Globally attracting invariant QLE manifolds
have recently been found to exist under certain parameter regimes in the continuous-time
two locus-two allele selection-recombination equations by \cite{baigentcompetitive}.

Similar analyses using quasi-equilibria involving variables other than linkage
disequilibrium have been conducted \citep{van1998unit, day2011bridging, lion2016spatial,
lion2018price}.  These models are evolutionary-ecological rather than population genetic
models, and rely on the weak selection approximation, but they still observe a rapid
convergence to quasi-linkage equilibrium.  Our approach to studying the QLE manifold is
very general, applicable to any system showing a significant separation of time-scales.
Any genetic system of this sort converges to quasi-linkage equilibrium and therefore under
an appropriate transformation of variables --- one which isolates the fast subsystem ---
can be analysed in a similar fashion.  Therefore, treating the QLE manifold as an slow
manifold and using linkage disequilibrium as a coordinate to approximate this surface
explicitly, is a powerful technique for other genetic systems and even evolutionary
ecological models.

Multi-locus models can have complex dynamics \citep{hastings1981stable, hofbauer1984hopf,
haig1991genetic, ubeda2019prdm9}. It appears that most analyses of multi-locus models have
been carried out under weak selection assumptions, in which case the dynamics are
relatively simple: stable cycling is generally not possible and the dynamics go to an
equilibrium \citep{nagylaki1999convergence}. The weak selection assumption allows for
general analytic results \citep{akin1982cycling, hofbauer1985selection, barton1995general,
nagylaki1999convergence, kirkpatrick2002general}, often invoking the use of the QLE.
Under weak selection, stable cycling and complex dynamics do not occur if the equilibria
are not degenerate and therefore complex dynamics are not observed under QLE.  This
association of QLE with weak selection and stability might have led to the impression that
complex dynamics are not compatible with quasi-linkage equilibrium
\citep{pomiankowski2004evolutionary}. What we have shown here is that complex dynamics are
possible and, furthermore, are played out in a state of \textit{quasi-linkage equilibrium}
showing the association between QLE and convergence to equilibrium to not be true in
general: it is possible to find continued fluctuations and sudden changes in the genetic
make up in a population at quasi-linkage equilibrium.}

\bibliographystyle{elsarticle-harv}
\bibliography{bibliography}

\appendix

\section{Deriving the discrete-time model} \label{sc:AppendixA}

\hl{Our model \citep{ubeda2019prdm9} can be written as a particular case of the model known as
the selection-recombination equations presented in \citep{lewontin1960evolutionary,
nagylaki1999convergence, burger2000mathematical, ubeda2005evolutionary} and many other
papers \citep{nagylaki1999convergence}. In the general model, haplotype frequencies evolve
according to
\begin{equation} \label{eq:generalmodel}
    \bar{w} x_i ' (n) = \sum_{j = 1}^m w_{i,j} x_i x_j
    + \epsilon_i \delta \left( w_{1,4} x_1 x_4 - w_{2,3} x_2 x_3 \right),
\end{equation}
where $x_i$ denotes the frequency of haplotype $i$, $m$ is the number of alleles and $n
\in \mathbb{N}_+$ represents the discrete time step. The recombination terms $\delta
\left( w_{1,4} x_1 x_4 - w_{2,3} x_2 x_3 \right)$ have different signs depending on the
haplotype, provided by $\epsilon_{i}$ for haplotype $i$. Specifically, for a two-locus
two-allele implementation of the model, $e_i$ is defined as}
\begin{equation}
    \epsilon_i=
    \begin{cases}
        -1 \quad &\text{for} \quad i = 1, 4\\
         1 \quad &\text{for} \quad i = 2, 3.\\
    \end{cases}
\end{equation}
The marginal mean fitness of a haplotype whose frequency is $x_i$ is given by
\begin{equation} \label{eq:generalMarginalfitness}
    w_i = \sum_{j=1}^n w_{i,j} x_j,
\end{equation}
and the mean fitness of the population is given by
\begin{equation} \label{eq:popmeanfitness}
    \bar{w} = \sum_{j=1}^n w_j x_j.
\end{equation}
Due to the normalisation of the right hand side of the governing equations of
the model by the mean fitness of the population, the sum of the haplotype
frequencies is always one. This means the state space of the model is the
simplex of dimension $n^m - 1$, where $n$ is the number of alleles and $m$ is
the number of loci.

\hl{Fitnesses for the two-locus two-allele version of our model are derived by computing all}
of the frequencies of offspring given by each possible mating combination.  Due to the
symmetries on the allele types determining when recombination occurs, the linkage
disequilibrium $D$ is the same for each haplotype and therefore can be taken out of the
fitness matrix. This is clearly true in the more general versions of the model, meaning
the linkage terms are separate in the statement of the general model equations
\eqref{eq:generalmodel}. After this, and other simplifications which are possible due to
\hl{symmetries in the gene conversion process and the viability benefits derived from
crossover, we arrive at the following fitness matrix for the two allele two loci version
of the model}
\begin{equation} \label{eq:fitnessmatrix}
    W = \left( \begin{array}{cccc}
        1 + \beta   & 1 - \gamma  & 1 & 1 \\
        1 + \gamma  & 1 -  \beta  & 1 & 1 \\
        1           & 1           & 1 - \beta   & 1 + \gamma \\
        1           & 1           & 1 - \gamma  & 1 + \beta
    \end{array} \right).
\end{equation}
Applying our specific fitness matrix to the general model given gives
the following system of equations
\begin{equation} \label{eq:discretesystem}
    \begin{aligned}
        \bar w x_1(n+1) &= (1 + \beta)x_1^2 + (1 - \gamma)x_1 x_2 + x_1 x_3
        + x_1 x_4 - \delta D,\\[5pt]
        \bar w x_2(n+1) &= (1 - \beta)x_2^2 + (1 + \gamma)x_2 x_1 + x_2 x_3
        + x_2 x_4 + \delta D,\\[5pt]
        \bar w x_3(n+1) &= (1 - \beta)x_3^2 + (1 + \gamma)x_3 x_4 + x_3 x_1
        + x_3 x_2 + \delta D,\\[5pt]
        \bar w x_4(n+1) &= (1 + \beta)x_4^2 + (1 - \gamma)x_4 x_3 + x_4 x_1
        + x_4 x_2 - \delta D.
    \end{aligned}
\end{equation}
Expanding the brackets in system \eqref{eq:discretesystem} and applying the conservation
law for the total population, $\sum_{i=1}^4 x_i = 1$, we can simply the system to
\begin{equation} \label{eq:discretesystem1}
    \begin{aligned}
        \bar w x_1(n+1) &= x_1(n) [1 + \beta x_1(n) - \gamma x_2(n)] - \delta D,\\
        \bar w x_2(n+1) &= x_2(n) [1 - \beta x_2(n) + \gamma x_1(n)] + \delta D,\\
        \bar w x_3(n+1) &= x_3(n) [1 - \beta x_3(n) + \gamma x_4(n)] + \delta D,\\
        \bar w x_4(n+1) &= x_4(n) [1 + \beta x_4(n) - \gamma x_3(n)] - \delta D,
    \end{aligned}
\end{equation}
where $\bar{w} \mathbf{x}(n+1)= \mathbf{f}(\mathbf{x})$ and $n \in \mathbb{N_+}$ and the
population mean fitness is
\begin{equation} \label{eq:popMeanFitnessOurs}
    \bar w = \sum_{i=1}^4 f_i (\mathbf{x}) =
    x_1 + x_2 + x_3 + x_4 + \beta (x_1^2 + x_4^2 - x_2^2 - x_3^2).
\end{equation}

\hl{\section{Isolation of the multiple time-scales} \label{sc:AppendixB}

The region of parameter space for which the following arguments hold is where the
heteroclinic cycle exists and is attracting in the discrete-time model, i.e. $\beta <
\gamma$.

\subsection{Time-scale separation nearby the interior equilibrium}

We find three distinct time-scales in the dynamics of the linearised system nearby the
interior equilibrium.  Recall that the eigenvalues of the interior equilibrium of the
continuous-time model are given by
\begin{equation}
    \begin{aligned}
        \lambda_1 &=  \frac{\gamma D^* + \sqrt{(\gamma D^*)^2 + \tfrac{1}{4}
        \beta (\beta - \gamma)}}{\bar w^*}, \\
        \lambda_2 &=  \frac{\gamma D^* - \sqrt{(\gamma D^*)^2 + \tfrac{1}{4}
        \beta (\beta - \gamma)}}{\bar w^*}, \\
        \lambda_3 &=  - \frac{\delta + 2D^*(\beta - \gamma)}{\bar w^*},
    \end{aligned}
\end{equation}
where $\bar w^*=1+ 2 \beta D^*.$ If $\beta>\gamma$ then $D^*>0.$ The interior equilibrium
in that case is a saddle. If $\beta<\gamma$ then $D^*<0$. Eigenvalues $\lambda_1$ and
$\lambda_2$ then are complex with negative real parts and the interior equilibrium is
always locally stable.

We introduce the parameter
\begin{equation}
    \label{eq:epsilon_definition}
    \epsilon = \sqrt{\gamma-\beta},
\end{equation}
which is small near the boundary of the region of parameter space in which we observe
time-scale separation, $\beta < \gamma$. We substitute this definition into the equations
and compute the eigenvalues at the interior equilibrium \eqref{eq:internalEquilibrium}.
For $0 < \epsilon \ll 1$, the eigenvalues satisfy the identities
\begin{equation}
    \begin{aligned}
        \bar w^*\lambda_1 &= -\epsilon^2\frac{\gamma}{8 \delta}  
        + i \epsilon \frac{ \sqrt{\gamma}}{2}+ O(\epsilon^3), \\
        \bar w^*\lambda_2 &=  -\epsilon^2\frac{\gamma}{8 \delta} 
        - i \epsilon \frac{ \sqrt{\gamma}}{2}+ O(\epsilon^3), \\
        \bar w^*\lambda_3 &=  - \delta +O(\epsilon^3).
    \end{aligned}
\end{equation}
The dynamics of the system linearised around the interior equilibrium
\eqref{eq:internalEquilibrium} operate on three distinct time-scales: $\bar w
\delta^{-1}$, $2 \bar w \epsilon^{-1} \gamma^{-\tfrac{1}{2}}$ and $8 \delta \bar w
\epsilon^{-2} \gamma^{-1}$. If $0<\epsilon \sqrt{\gamma}\ll 2\delta<1$ the time scales
separate as $\delta^{-1} \ll 2\epsilon^{-1} \gamma^{-\tfrac 12} \ll 2 \delta
\left(2\epsilon^{-1} \gamma^{-\tfrac 12}\right)^2.$ The second and third time-scales are
associated with the motion within the QLE manifold, while the first relates to the
approach towards the QLE manifold. Under this condition, the approach is very fast
compared to the dynamics on the manifold, which justifies making a quasi-steady state
assumption. This behaviour can be observed in Figure~\ref{fig:manifold_two_trajectories}
where the approach to QLE is very fast with associated time-scale $\bar w \delta^{-1}$,
and much faster than the cyclic behaviour on the manifold, which acts on time-scale $2
\bar w \epsilon^{-1} \gamma^{-\tfrac 12}$, which in turn is faster than the approach to
equilibrium which acts on time-scale $8 \delta \bar w \epsilon^{-2} \gamma^{-1}$.

Note that the separation of time-scales is a direct consequence of the double degeneracy
of the interior equilibrium \eqref{eq:internalEquilibrium}. Specifically, when $\beta =
\gamma$, and hence $\epsilon = 0$, two eigenvalues are zero. If the third eigenvalue is
much smaller than zero, for small $\epsilon$ and continuous dependence of the eigenvalues
on $\epsilon$, the separation of time scales follows. This implies that the existence of a
two-dimensional slow manifold is a generic result in the proximity of a double degeneracy
and independent of the details of the model.

\subsection{Time-scale separation in the full system}

We introduce the new variables
\begin{equation}
    \label{eq:XYZ_variables}
    \begin{aligned}
        X &= \sqrt{\gamma-\beta}\ln\biggl(\frac{A}{1-A}\biggr) + 
        \sqrt{\beta}\ln\biggl(\frac{B}{1-B}\biggr),\\
        Y &= (\gamma-\beta)\ln\bigl(A(1-A)\bigr)
        +\beta\ln\bigl(B(1-B)\bigr),\\
        Z &= \frac{D}{\gamma-\beta}.
    \end{aligned}
\end{equation}
If $\gamma \ne \beta$, these definitions implicitly define $A$ and $B$ locally as
functions of $X$ and $Y$ and therefore the inverse transformation exists.

Rewriting the continuous-time model \eqref{eq:continuousTransformedSystem} in the new
variables \eqref{eq:XYZ_variables},
\begin{equation}
    \begin{aligned}
        \frac{\df X}{\df t} &=
        \frac{\sqrt{\beta(\gamma-\beta)}}{\bar w}\left(\sqrt{\beta} (2B - 1) + \sqrt{\gamma - \beta}
        (2A - 1) + \frac{\gamma \sqrt{\gamma-\beta} (2 B - 1) Z}{B(1-B)}\right),\\
        \frac{\df Y}{\df t} &= -\frac {\beta(\gamma-\beta)}{\bar w} \gamma  \frac{ (1-2B)^2}{B(1-B)} Z,\\
        \frac{\df Z}{\df t} &= \frac{(\gamma-\beta)^{-1}}{\bar{w}}
        \begin{aligned}[t]
            \biggl[
                &(\gamma-\beta) \Bigl[(\gamma-\beta)^{2} Z^2-A B (1-A)(1-B)\Bigr]\\
                &\quad - (\gamma-\beta)Z (\beta (2 A - 1)(2 B - 1) + \delta )
            \biggr].\\
        \end{aligned}
    \end{aligned}
\end{equation}
Using \eqref{eq:epsilon_definition}, this can be written as
\begin{equation} \label{eq:systemDistinctTime}
    \begin{aligned} 
        \frac{1}{\epsilon} \frac{\df X}{\df t} &=
        \frac{\sqrt{\beta}}{\bar w}\left(\sqrt{\beta} (2B - 1) + \epsilon
        (2A - 1)+\frac{\gamma \epsilon (2 B - 1) }{B(1-B)}Z\right),\\
        \frac{1}{\epsilon^2} \frac{\df Y}{\df t} &=
        -\frac {\beta \gamma}{\bar w} \frac{ (1-2B)^2}{B(1-B)} Z,\\
        \frac{\df Z}{\df t} &= \frac{1}{\bar{w}}
        \biggl[ 
            \epsilon^{4} Z^2-A B (1-A)(1-B) -
            Z \left(\beta(2 A - 1)(2 B - 1) + \delta \right)
        \biggr].
    \end{aligned}
\end{equation}
When $\epsilon$ is small, the form of \eqref{eq:systemDistinctTime} isolates three
distinct time-scales. The variable $Z$ is changing at the fastest time-scale, and for $Z$
small the variables $X$ and $Y$ (and $A$ and $B$) are effectively constant. If $A$ and $B$
are constant, the variable $Z$ has an equilibrium at
\begin{equation}
    \begin{aligned}
         Z^*= &\epsilon^{-2}
         \frac{\beta (2 A - 1)(2 B - 1) + \delta }
         {2\epsilon^2} \\ & - \epsilon^{-2} \sqrt{\left(\frac{\beta
         (2 A - 1)(2 B - 1) + \delta}{2\epsilon^2}\right)^2 + 
         A B (1-A)(1-B)}.
    \end{aligned}
\end{equation}
The linearised dynamics around $Z^*$ are given by         
\begin{equation}
    \frac{\df (Z - Z^*)}{\df t}=-\frac{1}{\bar{w}} (Z - Z^*)
    \sqrt{\biggl(\beta (2 A - 1)(2 B - 1) + \delta\biggr)^2 
    + 4 \epsilon^4 A B (1-A)(1-B)}
\end{equation}
which always converges to the equilibrium $Z = Z^*$. Based on this we choose
$D_{QLE}=\epsilon^2 Z^*$. If $D_{QLE}$ is situated outside the simplex this argument is
not relevant but a similar argument can be applied for attraction to the state
$Z=\epsilon^{-2} D_S.$

\section{Determining the eigenvalues of the corner equilibria} \label{sc:AppendixC}}

In the vicinity of the origin ($\Phi_4$), we find by Taylor expanding to second order
that the QLE manifold is approximately defined by $D_{QLE}(0,0)\approx -\frac{\gamma-\beta
}{\beta +\delta }A B$. The attracting manifold $D=D_S(A,B)$ in the vicinity of the origin
is approximately
\begin{equation}
    D_S(A,B) \approx \biggl\{
    \begin{split}
        -A B         \quad &\text{if} \quad  \delta \le \gamma-2 \beta, \\
        D_{QLE}(A,B) \quad &\text{if} \quad \delta > \gamma-2 \beta.
    \end{split}
\end{equation}
We then find for the eigenvalues
\begin{table}[h]
    \centering
    \begin{tabular}{c | c | c | c | c | c }
        Eigenvalue & $1-\frac{\beta }{1+\beta}$ &
        $1+\frac{\gamma -\beta }{1+\beta }$  & $1-\frac{\beta }{1+\beta }$ &
        $1+\frac{\gamma -\beta }{1+\beta }$  \\[6pt] \hline
        Type & $c_j$ & $e_j$ & $ c_j $ & $ e_j $ \\[6pt] \hline
        Condition & \multicolumn{2}{c}{$\delta \le\gamma-2 \beta$} \vline &
        \multicolumn{2}{c}{$\delta >\gamma-2 \beta$}
    \end{tabular}
    \caption{The eigenvalues of the equilibria $\Phi_1$ and $\Phi_4$. The
    eigenvalues do not depend on the condition.}
\end{table}

Likewise, in the vicinity of the equilibrium $\Phi_2$ and $\Phi_3$ the QLE manifold
is approximately
\begin{equation}
    D_{QLE}(A,B) \approx \left\{
    \begin{array}{cc}
        - \frac{\beta-\delta}{\gamma -\beta}+\frac{2 \beta }{\gamma -\beta }A+
        \frac{2 \beta }{\gamma-\beta }(1-B)+ (\frac{\gamma -\beta }{\beta -\delta }+
        \frac{4 \beta }{\gamma -\beta })A(B-1) &\mbox{if } \delta<\beta,\\
        -\frac{\gamma-\beta }{\delta-\beta }A (1-B) &\mbox{if } \delta>\beta,
    \end{array}\right.
\end{equation}
and hence
\begin{equation}
    D_S(A,B) \approx \left\{
    \begin{array}{cc}
        \max( -A B,-(1-A)(1-B))& \mbox{if }\delta \le \beta, \\
        D_{QLE}(A,B) &\mbox{if }\delta>\beta,
    \end{array}\right.
\end{equation}
We then find for the eigenvalues
\begin{table}[h]
    \centering
    \begin{tabular}{c | c | c | c | c | c }
        Eigenvalue & $1-\frac{\gamma-\beta }{1-\beta}$ &
        $1+\frac{\beta }{1-\beta}$   & $1-\frac{\gamma-\beta }{1-\beta}$ &
        $1+\frac{\beta }{1-\beta}$  \\[6pt] \hline
        Type & $c_j$ & $e_j$ & $ c_j $ & $ e_j $ \\[6pt] \hline
        Condition & \multicolumn{2}{c}{$\delta \le \beta$} \vline &
        \multicolumn{2}{c}{$\delta >\beta$}
    \end{tabular}
    \caption{The eigenvalues of the equilibria $\Phi_2$ and $\Phi_3$. The eigenvalues do
    not depend on the condition}
\end{table}

\end{document}